\documentclass[aps,twocolumn,showpacs,preprintnumbers,amsmath,amssymb,footinbib]{revtex4}
\usepackage{epsf}
\usepackage{hyperref}
\newcommand{\beq}{\begin{equation}}
\newcommand{\boldl}{{\bf L}}
\newcommand{\calc}{{\cal C}}
\newcommand{\calp}{{\cal P}}

\newcommand{\calv}{{\cal V}}
\newcommand{\calvc}{{\cal V}_{cons.}}
\newcommand{\calz}{{\cal Z}}
\newcommand{\dslash}{D\!\!\!\!\slash}
\newcommand{\detrm}{{\rm det}}
\newcommand{\eeq}{\end{equation}}

\newcommand{\ellf}{\ell_N}
\newcommand{\ellaf}{\ell_{\overline{N}}}
\newcommand{\ellt}{\ell_3}
\newcommand{\ellat}{\ell_{\overline{3}}}

\newcommand{\real}{{\rm Re}}
\newcommand{\imag}{{\rm Im}}
\newcommand{\rme}{{\rm e}}
\newcommand{\tr}{{\rm tr}}

\newcommand{\veff}{{\cal V}_{e\!f\!f}}
\newcommand{\veig}{{\cal V}_{V\!dm}}

\newcommand{\zn}{$Z(N)\;$}

\def\cqg#1#2#3{Class. Quant. Grav. {\bf #1}, #2 (#3)}

\def\ibid#1#2#3{{\it ibid.} {\bf #1}, #2 (#3)}

\def\jhep#1#2#3{Jour. of High Energy Phys. {\bf #1}, #2 (#3)}

\def\npb#1#2#3{Nucl. Phys. B {\bf #1}, #2 (#3)}
\def\npps#1#2#3{Nucl. Phys. Proc. Suppl. {\bf #1}, #2 (#3)}
\def\npsb#1#2#3{Nucl. Phys. Proc. Suppl. B {\bf #1}, #2 (#3)}
\def\plb#1#2#3{Phys. Lett. B {\bf #1}, #2 (#3)}
\def\prc#1#2#3{Phys. Rev. C {\bf #1}, #2 (#3)}
\def\prd#1#2#3{Phys. Rev. D {\bf #1}, #2 (#3)}
\def\prl#1#2#3{Phys. Rev. Lett. {\bf #1}, #2 (#3)}

\def\ptp#1#2#3{Prog. Theor. Phys. {\bf #1}, #2 (#3)}
\def\ptps#1#2#3{Prog. Theor. Phys. Suppl. {\bf #1}, #2 (#3)}

\begin{document}
\title{Dense Quarks, and the Fermion Sign Problem, in a $SU(N)$ Matrix
Model}
\author{Adrian Dumitru,$^{a}$ Robert D. Pisarski,$^{b}$
and Detlef Zschiesche$^{a}$}
\affiliation{
$^a$Institut f\"ur Theoretische Physik, J.~W.~Goethe Univ.,
Max-von-Laue-Strasse 1,
D-60438 Frankfurt am Main, Germany\\
$^b$Nuclear Theory Group,
Brookhaven National Laboratory, Upton, NY, 11973, U.S.A.\\
}
\begin{abstract}
We study the effect of dense quarks in a $SU(N)$ matrix model
of deconfinement.  For three or more colors, the quark contribution
to the loop potential is complex.  After adding the charge conjugate
loop, the measure of the matrix integral is real, but not
positive definite.  In a matrix model, quarks 
act like a background $Z(N)$ field;
at nonzero density, the background field also has an imaginary part, 
proportional to the imaginary part of the loop.  Consequently, while
the expectation values of the loop and its complex conjugate are both
real, they are not equal.  These results
suggest a possible approach to the fermion sign problem in lattice QCD.
\end{abstract}
\date{\today}
\maketitle

At nonzero temperature, numerical simulations in lattice QCD
have provided fundamental insight into
the transition from a hadronic, to a deconfined, chirally symmetric 
plasma \cite{lattice_review_temp}.  At nonzero quark density,
however, at present simulations 
are stymied by the ``fermion sign problem'' 
\cite{lattice_review_density,hasenfratz_karsch,karsch_wyld,gibbs,gocksch,glasgow,stephanov,akw,blum,deforcrand_lal,langfeld_shin,potts1,potts2,AKKS,merons_etc,factorization}.  
Even in the limit of high temperature, and small chemical potential,
only approximate methods can be used 
\cite{fodor_katz,swansea_bielefeld,deforcrand_philipsen,gavai_gupta,splittorff1,splittorff2}.

In this paper we consider deconfinement in a mean field approximation
to a model of thermal Wilson 
lines \cite{polyakov_loop_old,mclerran_svetitsky}, 
which is a matrix model 
\cite{banks_ukawa,matrix_models_density,polyakov_loop,DRR,sannino,fukushima,dhlop,dlp,bielefeld_loop,aharony,schnitzer,aharony2}.  In
Sec.~\ref{suN_matrix_model} we discuss general features of $SU(N)$
matrix models at nonzero quark density \cite{matrix_models_density}.  
In sec.~\ref{U1_matrix_model},
this is briefly contrasted with the (trivial) case
of a $U(1)$ model \cite{gibbs}.
Numerical results for three colors are presented 
in Sec.~\ref{N=3_matrix_model}.
In Sec.~\ref{Lattice_QCD}, we conclude with some remarks about
some methods which might be of use for dense quarks in lattice QCD.

\section{$SU(N)$ Matrix Model}
\label{suN_matrix_model}

In a gauge theory at nonzero temperature, a basic
quantity is the thermal Wilson line, 
$\boldl = \calp \exp(i g \int A_0 \, d\tau)$, where $g$ is the gauge
coupling, $A_0$ is the timelike component of the vector potential,
and the integral over the imaginary time, $\tau$, runs from 
$0$ to $1/T$, where $T$ is the temperature \cite{polyakov_loop_old}.  
An effective theory of thermal Wilson lines, interacting with
static magnetic fields, can be constructed, and is valid in
describing correlations over spatial distances $\gg 1/T$ 
\cite{banks_ukawa,polyakov_loop,DRR,sannino,fukushima,dhlop,dlp,aharony,schnitzer,aharony2}.

Over large distances, we use a mean field approximation
to this effective theory.  This gives 
an integral over a single Wilson line, $\boldl$,
with the partition function that of a matrix model:
\beq
\calz = \int \; d\boldl \; \exp \left( - (N^2 - 1 ) \left(
\calv_{gl}(\boldl) + \calv_{qk}(\boldl) \right) \right) \; ;
\label{partition_function}
\eeq
$\boldl$ is an $SU(N)$ matrix, satisfying
$\boldl^\dagger \boldl = {\bold 1}$ and ${\rm det}\; \boldl = 1$.
Under gauge transformations $\Omega$, it transforms as
$\boldl \rightarrow \Omega^\dagger \boldl \Omega$, so that
gauge invariant quantities are formed by taking 
traces of $\boldl$.  These are Polyakov loops.  In the matrix model,
the effects of gluons and quarks are represented by potentials,
$\calv_{gl}(\boldl)$ and $\calv_{qk}(\boldl)$, 
which are (gauge invariant) functions of the Wilson line.  
The effects of fluctuations, which are not included in the matrix
model, can also be included in a systematic fashion
\cite{dlp,oswald_pisarski}.

The pure glue theory is invariant under a global symmetry of \zn,
and so this must be a symmetry of
the gluon loop potential, ${\cal V}_{gl}(\boldl)$.
The simplest form for the gluon loop potential is a type
of mass term,
\beq
\calv_{gl}(\boldl) = -\, m^2 \; \ellaf \; \ellf \;\;\; , \;\;\;
\ellf = \frac{1}{N} \; \tr \; \boldl \; ,
\label{gluon_potential}
\eeq
where $\ellf$, and
$\ellaf = \ellf^*$, are the Polyakov loops in the
fundamental, and anti-fundamental, representations.  Up to a constant, 
this gluon potential is proportional to the Polyakov
loop in the adjoint representation.

In general, the gluon potential is a sum over all loops in \zn neutral
representations \cite{dlp}; this can be written as a power series
in terms like $(|\ellf|^2)^2$, {\it etc.}
These terms are invariant under a larger
global symmetry of $U(1)$.  The first term which is invariant under
\zn, but not $U(1)$, is
\beq
(\ellf)^N + (\ellaf)^N \; .
\label{zna}
\eeq
Another such term is
\beq
i \left( (\ellf)^N - (\ellaf)^N \right) \; ,
\label{znb}
\eeq
where the factor of $i$ is added to ensure that in all, the term is real.

While (\ref{zna}) certainly appears in the gluon loop potential,
terms such as (\ref{znb}) should not
arise in effective theories of relevance to QCD.  Gluons are
invariant under the discrete symmetry of
charge conjugation, $\calc$, under which 
$A_\mu \rightarrow - A_\mu^*$ (taking $A_\mu = A_\mu^a t^a$, 
and Hermitean generators $t^a$ for $SU(N)$) \cite{zinn_justin}.  
Under $\calc$, 
the Wilson line transforms into its complex conjugate,
$\boldl \rightarrow \boldl^*$, so that 
(\ref{zna}) is even under $\calc$, and (\ref{znb}), odd.

Quarks in the fundamental representation of $SU(N)$
are not invariant under the global \zn symmetry.  
Thus quarks tend to induce a background \zn magnetic field,
which we characterize by a parameter $h$.
The simplest contribution to the quark loop potential is then
\cite{banks_ukawa,matrix_models_density,polyakov_loop,sannino,fukushima}
\beq
\calv_{qk}(\boldl) = - \frac{h}{2} \left( {\rm e}^{\mu} \; \ellf \; +
\; {\rm e}^{-\mu} \; \ellaf \right) \; .
\label{quark_potential}
\eeq

At finite $N$, $h \neq 0$ affects the deconfining transition in
the standard manner of a background magnetic field
\cite{banks_ukawa,sannino}.
If the deconfining transition is of first order
in the absence of quarks, then their presence tends
to weaken the transition.  Eventually, it disappears at a
critical end-point, for some value of $h$; above this value,
there is no phase transition, just a smooth crossover.  If the deconfining
transition is of second order in the absence of quarks, then any
background field, $h \neq 0$, washes out the transition.

At infinite $N$, if one is away from the Gross--Witten point, then
the behavior is like that at finite $N$.  
Precisely at
the Gross--Witten point \cite{dhlop,dlp}, correlation
lengths diverge at the transition.
Then like a second order transition, any background field
changes the order: from
first order at the Gross--Witten point, 
into one of third order when $h \neq 0$ \cite{dlp,aharony}.

We have also added a parameter, $\mu$, to 
represent the quark chemical potential; $\mu$ 
should be understood as the true quark chemical
potential, divided by temperature.  The quark chemical potential is
associated with a conserved charge for the global $U(1)$ symmetry of
baryon number.  This dictates that the chemical potential enters
in the above form, like the imaginary component of a $U(1)$ gauge field
\cite{hasenfratz_karsch}.    

Under charge conjugation, the Wilson line transforms into its
complex conjugate, and the chemical potential changes sign:
\beq
\calc: \;\;\;
\boldl \rightarrow \boldl^* \;\;\; , \;\;\; \mu \rightarrow - \mu \; .
\label{charge_conjugation}
\eeq
The term in (\ref{quark_potential}) is invariant under $\calc$,
as should be all terms in the quark loop potential.

Implicitly, we have integrated out the quarks to obtain the 
loop potential in (\ref{quark_potential}).  For example, if one
computes the quark determinant in a background gauge field,
$\sim \tr\, \log (\dslash + m_{qk})$, one will obtain a term such as 
(\ref{quark_potential}): see, for example, the calculations
of Langfeld and Shin \cite{langfeld_shin} and of Schnitzer \cite{schnitzer}.
Other discussions of loop potentials with quarks include
those of \cite{polyakov_loop,DRR,sannino,fukushima};
at nonzero quark density, see \cite{matrix_models_density,fukushima}.
These calculations show that at a temperature $T$,
the background field of massive quarks behaves as
$h \sim \exp(- m_{qk}/T)$, 
reaching some finite value as the quark mass vanishes.
As with the gluon loop potential in (\ref{gluon_potential}),
there are many other terms besides that of (\ref{quark_potential})
in $\calv_{qk}$.
These involve all possible traces of
\beq
{\rm e}^\mu \; \boldl \; \;\; {\rm and} \;\;\;
{\rm e}^{- \mu} \; \boldl^* \; .
\eeq
in such combinations which are
invariant under charge conjugation, (\ref{charge_conjugation}).  
These two matrices represent, respectively,
the propagation of a particle forward in 
imaginary time, and an anti-particle backward in time.
Of course, charge conjugation symmetry is violated by
a given value of $\mu \neq 0$: $\calc$ just implies that
a Fermi sea of quarks behaves similarly to one of anti-quarks
(neglecting electro-weak interactions).

The quark contribution to the loop potential equals
\beq
\calv_{qk}(\boldl) = - h \left( \cosh(\mu) \; \real \; \ellf \; +
\; i \, \sinh(\mu) \; \imag \; \ellf \right) \; ,
\label{qpotentialB}
\eeq
where $\real$ and $\imag$ denote the real 
and imaginary parts, respectively.  
At zero chemical potential, quarks generate a real background \zn field
for the real component of the loop, $\sim \real \, \ellf$.
When the
chemical potential is nonzero, however, the background \zn field 
not only contains a piece proportional to the imaginary part of the
loop, $\sim \imag \, \ellf$, but with a coefficient which is itself
imaginary.  

The case of two colors is special.  
For two colors, loops in any representation are real, and 
for any $\mu$, the 
background field generated by quarks is always real.
For three or more colors, however, loops have imaginary parts,
and the potential generated by quarks is manifestly complex,
(\ref{qpotentialB}).  This is how the fermion sign problem
appears in a matrix model.

In this case, though, it is easy to reduce the sign problem,
which appears to be one of complex phases, to one in which the
phases are always real.  If a given matrix, 
$\boldl$, contributes to the partition function, then so does
its charge conjugate, $\boldl^*$.  Adding the contributions of
$\boldl$ and $\boldl^*$ together, we can rewrite the partition
function in a form which is manifestly real,
\beq
\calz = \int \; d\boldl \; {\rm e}^{- \widetilde{\calv}(\boldl) }
\cos ( \widetilde{h} \; \imag \; \ellf ) 
\label{partition_func_real}
\; ,
\eeq
where
\begin{eqnarray}
\widetilde{\calv}(\boldl) &=&
(N^2 - 1) \left(
\calv_{gl}(\boldl) - h \; \cosh(\mu) \; \real \; \ellf \right) \; , \;
\label{Vtilde}\\
%\nonumber\\
\widetilde{h} &=& (N^2 - 1) \, h \, \sinh(\mu) \; .
\end{eqnarray}
The potential $\widetilde{\calv}(\boldl)$ is even under charge
conjugation of the gluons, while $\widetilde{h} \; \imag\, \ellf$ is
odd.  We can use this to write the expectation value of the fundamental
loop as
\begin{eqnarray}
&&\langle \ellf \rangle
= \frac{1}{\calz} \; \int d\boldl \; \;
{\rm e}^{- \widetilde{\calv}(\boldl) } \times
\nonumber\\
&& \;\;\;
\left( \cos ( \widetilde{h} \, \imag \, \ellf ) \, \real \, \ellf
 - \sin ( \widetilde{h} \, \imag \, \ellf )\, \imag \, \ellf \right) ,
\label{vev_loop}
\end{eqnarray}
while that of the charge conjugate loop is
\begin{eqnarray}
&&\langle \ellaf \rangle
= \frac{1}{\calz} \; \int d\boldl \;\; 
{\rm e}^{- \widetilde{\calv}(\boldl) } \times
\nonumber\\
&& \;\;\;
\left( \cos ( \widetilde{h} \, \imag \, \ellf ) \, \real \, \ellf
 + \sin ( \widetilde{h} \, \imag \, \ellf )\, \imag \, \ellf \right) .
\label{vev_conj_loop}
\end{eqnarray}
Because dense quarks induce an imaginary background field 
for the imaginary part of $\ellf$, the expectation values
of $\ellf$ and $\ellaf$ are {\it not} equal to one another, although
they are both real.  

Physically, this is natural.  A loop
is proportional to the (trace of the) wave function of a quark; 
the complex conjugate loop, to that of 
an anti-quark.  A Fermi sea represents a net excess of
quarks over anti-quarks, so at $\mu \neq 0$, 
quarks and anti-quarks propagate differently.
In a matrix model, this manifests itself as
unequal expectation values for $\ellf$ and $\ellaf$.

Karsch and Wyld performed numerical simulations
for a model of $SU(3)$ matrices, living 
on sites of a three-dimensional lattice, at nonzero density \cite{karsch_wyld}.
Our matrix model represents a mean field approximation to their theory.
They were the first to observe
that the expectation values of $\ellf$ and $\ellaf$
differ at nonzero density.  This also happens for 
a Potts model at nonzero density \cite{potts1,potts2}.

This contrasts with what would happen if the background
field which coupled to the imaginary part of $\boldl$ was real; 
{\it i.e.}, for $\mu = i \widetilde{\mu}$.  This corresponds 
to a $U(1)$ rotation of $\boldl$, 
so that both expectation values are complex, and satisfy
$\langle \ellf \rangle = (\langle \ellaf \rangle)^*$.
In a $U(N)$ theory, this just rotates the vacuum by an angle
$= \widetilde{\mu}$; for 
$SU(N)$, because of the \zn symmetry, the vacuum structure
is more involved.

In Sec.~\ref{N=3_matrix_model} we present numerical calculations of
the expectation values of the fundamental and anti-fundamental loops
for $N=3$.  Even without explicit calculation,
however, we can understand the qualitative nature of the solutions.

Consider first the limit about zero chemical potential.
Taking the derivatives of the expectation values in (\ref{vev_loop})
and (\ref{vev_conj_loop}) with respect to $\mu$, 
\beq
\frac{\partial \langle \ellf\rangle}{\partial \mu}\biggr|_{\mu = 0} 
= -\; \frac{\partial \langle \ellaf\rangle}{\partial \mu}\biggr|_{\mu = 0} 
= - \; h (N^2-1) \langle (\imag \; \ell)^2 \rangle \Bigr|_{\mu = 0} \; .
\label{derivatives}
\eeq
About $\mu =0$, then, as $\mu$ increases, so does
$\langle \ellaf\rangle$, 
while $\langle \ellf\rangle$ decreases.  

It is also easy to understand the behavior of the expectation values
in the limit of large $\mu$.
This corresponds to a very strong background field, proportional
to $\sim \ellf$.  Taking 
the Wilson line $\boldl = \exp(i \omega)$, the real part
of $\ellf$ is $\sim \tr \; \omega^2$, while the imaginary part
is $\sim \tr \; \omega^3$.  For large background field, then,
the potential is dominated by the real part,
$\sim h \exp(\mu) \tr \; \omega^2$; fluctuations in the imaginary part are
suppressed, by $\sim \exp(-\mu/2)$ relative to the real part.
Thus as $\mu \rightarrow \infty$, the expectation values of
$\ellf$ and $\ellaf$ both approach unity,
$\langle \ellf \rangle\approx \langle \ellaf \rangle \rightarrow 1$.

(The parameter $\mu$ 
is the quark chemical potential divided by temperature, 
so naively, $T \rightarrow 0$ corresponds to 
$\mu \rightarrow \infty$.  Remember, though,
that our effective theory is only valid for distances $\gg 1/T$.
We believe that the large $\mu$ behavior is an artifact of the model,
and is not indicative what happens in 
the full theory at low temperature.  See, also, Sec. IV.)

We can thus anticipate the behavior of the expectation values of the
loops as a function of $\mu$. 
Due to the background \zn field of the quarks, both
loops are equal at $\mu = 0$.
As $\mu$ increases, at first the two expectation values 
split: one increases, while the other decreases.
As $\mu \rightarrow \infty$, they come together and approach unity.
For $N=3$, this is illustrated in Sec.~\ref{N=3_matrix_model}
by Fig.~\ref{fig:vevs_m2=0}.

It is customary to interpret the expectation
value of the Polyakov loop as the ``free energy'' of a test quark
\cite{mclerran_svetitsky}.
At nonzero density, this implies that the expectation value of the
fundamental loop is the ``free energy'' 
of a test quark, and that of the conjugate loop
is the ``free energy'' of an anti-quark \cite{karsch_wyld}:
\beq
\langle \ellf \rangle = \exp(-F_q/T) \;\;\; , \;\;\;
\langle \ellaf \rangle = \exp(-
F_{\overline{q}}/T) \; .
\label{expectations}
\eeq
Any free energy, however, should decrease monotonically 
with $\mu$; because $\langle \ellf \rangle$ decreases about
$\mu =0$, though, the ``free energy'' for a test quark increases with $\mu$.  
This quandry is resolved by recognizing that the expectation
values of the loops are not free energies, but just the traces
of test propagators \cite{polyakov_loop,dhlop}.  As such, they need not
behave monotonically with $\mu$.

\section{$U(1)$ Matrix Model}
\label{U1_matrix_model}

Before going into numerical results for $SU(3)$, 
we briefly discuss what happens in a $U(1)$ model,
as first proposed by Gibbs \cite{gibbs}.

For $U(1)$, the loop is just $\ell = \exp(i \theta)$, where
$\theta$ runs from $- \pi$ to $\pi$.  At a nonzero density $\mu$, we
take the partition function as
\beq
\calz = \int^{+\pi}_{-\pi} d\theta \;
\exp\left( \frac{h}{2} 
\left( \rme^{\mu} \ell + \rme^{-\mu} \ell^* \right) \right) \; .
\label{u1_part_fnc}
\eeq
Like $SU(N)$, the fermion contribution to the loop potential is
complex at nonzero chemical potential.  
Summing over a given $\theta$, plus its charge conjugate,
which is just $- \theta$, the partition function becomes
\beq
\calz = \int^{+\pi}_{-\pi} d\theta \;
\rme^{h \cosh(\mu) \cos\theta }
\cos\left( h \sinh(\mu) \sin\theta \right) \; ,
\eeq
which is real.  

As for $SU(N)$, the expectation value of a loop, and its charge
conjugate, are unequal when $\mu \neq 0$.  However,
in the original integral, (\ref{u1_part_fnc}), we can shift the
integration by
\beq
\theta \rightarrow \theta + i \mu \; .
\label{shift}
\eeq
Doing so, we find that the partition function is completely independent
of $\mu$.  In terms of expectation values, this implies that 
\beq
\rme^{\mu} \langle \; \ell \; \rangle_{\mu \neq 0}
\; = \; \rme^{- \mu} \langle \; \ell^* \; \rangle_{\mu \neq 0} 
\; = \; \langle \; \ell\; \rangle_{\mu = 0} \; .
\label{equality}
\eeq
This is an immediate consequence of the change in variables possible
in a $U(1)$ model, (\ref{shift}).

For $SU(N)$ loops, we found that both loops approach unity at
large $\mu$.  This is not true for $U(1)$ loops, (\ref{equality}):
as $\mu \rightarrow \infty$, 
$\langle \; \ell \; \rangle$ is very small, while
$\langle \; \ell^* \; \rangle$ is large.
The difference arises because for 
$U(1)$, the real part of the loop is $\cos\theta$,
while the imaginary part is $\sin \theta$.  At large $\mu$,
the real part is $\sim 1 - \theta^2/2 + \ldots$, while
the imaginary part is $\sim \theta$.  At large $\mu$, then,
for $U(1)$ the imaginary part of the loop dominates, instead of the real
part, as for $SU(N)$.

There is a simple physical reason why, for $U(1)$, the partition function
is independent of the fermion chemical potential \cite{gibbs}.
With a 
$U(1)$ gauge field, there is no way of forming
baryons: the only states which
are neutral under $U(1)$ are trivial,
having an equal number of fermions,
$\exp(i\theta)$, and anti-fermions, $\exp(- i\theta)$.

There is a less trivial consequence of this observation.  Consider
a $SU(N)$ gauge theory.  Up to corrections
$\sim 1/N$, at large $N$, there is no difference
between the measure for $SU(N)$ and that for $U(N)$.
For a $U(N)$ gauge theory, however,
we can rotate the quark chemical potential away.  To the extent
that $SU(N)$ is close to $U(N)$, then, 
at large $N$ the effects of quark chemical
potential appear only in terms which are subleading.

Put more directly, assume that deconfinement, and chiral symmetry
restoration, occurs at some temperature $T_d \approx T_\chi$ when
$\mu = 0$.  Then the natural scale at which the quark chemical
potential matters is {\it larger} than $T_d$ by some 
(fractional) power of $N$, which can be computed in a matrix model
\cite{oswald_pisarski}.  

\section{$N=3$ Matrix Model}
\label{N=3_matrix_model}
In this section we present numerical results for three colors,
where the $SU(3)$ matrix model is just a two dimensional
integral.  When the chemical potential $\mu$ is real,
and the background field $h$ is large, 
the integrands
of~(\ref{partition_func_real}), (\ref{vev_loop}), and (\ref{vev_conj_loop})
oscillate strongly. Nevertheless, 
we show that the value of these integrals are {\em not}
sensitive to large cancellations of positive and negative
contributions, and can
be computed numerically without great difficulty.

For three colors, the loop potential is a function
of the triplet and anti-triplet loops,
\beq
\ellt \equiv \frac{1}{3} \;\tr\;\boldl\quad,\quad
\ellat \equiv \frac{1}{3} \;\tr\;\boldl^\dagger~.
\eeq
We straightforwardly extend the analysis of \cite{dlp},
going into some detail in order to avoid confusion.
Previously, we assumed that the expectation values of the 
triplet and anti-triplet loops are equal; now we must allow
that they can differ.  
In the partition function of (\ref{partition_function}), 
we introduce two fields, $\lambda$ and $\lambda_*$, which are
the values of these loops for a given matrix, $\boldl$:
\begin{eqnarray}
\calz &=& \int d\boldl \int d\lambda_* \int d\lambda \;
\delta(\lambda_* - \ellat) \; 
\delta(\lambda - \ellt) \;
\nonumber\\
\; &&\exp\left( - 8 ( \calv_{gl}(\lambda_* \lambda) 
+\calv_{qk}(\lambda_*,\lambda) )\right) \; .
\label{partition_function_constrained}
\end{eqnarray}
We then exponentiate the constraints by introducing fields
$\overline{\omega}_*$ and ${\overline \omega}$,
$$
\calz = \int d\lambda_* \int d\overline{\omega}_*  
\int d\lambda \int d\overline{\omega} 
\int d\boldl \;
\exp\left( - 8 \calvc \right) \; ,
$$
\beq
\calvc =
\calv_{gl} + \calv_{qk}
+ i \overline{\omega}_* (\lambda_*  - \ellat) 
+ i \overline{\omega} (\lambda  - \ellt) \; .
\label{constrained_pot}
\eeq
At all stationary points
$i \overline{\omega}$ and $i \overline{\omega}_*$ are real, so
we define $\omega = i \overline{\omega}$ and 
$\omega_* = i \overline{\omega}_*$.

Next, we define the matrix integral
\beq
\calz_{GW}(\omega_*,\omega) = \int d\boldl \; \exp\left( \; 8 \, 
\left( \omega_* \; \ellat  + \omega \; \ellt  \right)\right) \; .
\label{gross_witten_integral}
\eeq
For given values of $\omega_*$ and $\omega$, the expectation
values of the fields are 
\begin{eqnarray}
\ell^0_*(\omega_*,\omega) &=& 
 \frac{1}{\calz} \int d\boldl \; \ellat \;
   \exp\left( 8
\left(\omega\; \ellt +\omega_* \; \ellat \right)\right)~,
\nonumber\\
\ell^0(\omega_*,\omega) &=& \frac{1}{\calz} \int d\boldl \; \ellt\;
   \exp\left( 8
\left(\omega\; \ellt +\omega_* \; \ellat \right) \right)~.
\label{lambdas} 
\end{eqnarray}
We introduce the Vandermonde potential, as a function of two fields,
$\ell$ and $\ell_*$, through Legendre transformation,
\begin{eqnarray}
\calz_{GW}(\omega_*,\omega)
&=& \int d\ell_* \int d\ell \; 
\nonumber\\
&& \hspace{-1.0cm}\times \exp(
8 ( \omega_* \ell_* + \omega \ell - 
\veig(\ell_*,\ell) )) .
\label{legendre}
\end{eqnarray}
The stationary point of this integral is for
\begin{eqnarray}
\omega_*(\ell^0_*,\ell^0) &=& \left.
\frac{\partial \veig(\ell_*,\ell)}{\partial \ell_*}
\right|_{(\ell_* = \ell^0_*\, ,\, \ell = \ell^0)} \; , 
\nonumber \\
\omega(\ell^0_*,\ell^0) &=& \left.
\frac{\partial \veig(\ell_*,\ell)}{\partial \ell}
\right|_{(\ell_* = \ell^0_*\, , \, \ell = \ell^0)} \; .
\label{stationary_leg}
\end{eqnarray}
This satisfies the consistency condition
\beq
\frac{\partial\omega_*(\ell^0_*,\ell^0)}{\partial\ell^0} =
\frac{\partial\omega(\ell^0_*,\ell^0)}{\partial\ell_*^0}~.
\label{consistency}
\eeq

For given values of $\omega_*$ and $\omega$, we numerically
computed the integrals in (\ref{lambdas}), to obtain $\ell_*^0$
and $\ell^0$.  We then invert them, to obtain $\omega_*$ and
$\omega$, as a function of $\ell^0_*$ and $\ell^0$. The Vandermonde
potential then follows:
\beq
\veig(\ell_*,\ell) = \int\limits_0^\ell d\ell^0
\; \omega(0,\ell^0) + 
\int\limits_0^{\ell_*} d\ell_*^0
\; \omega_*(\ell_*^0,\ell) \; .
\eeq
We have chosen a definite path to go from $(0,0)$ to $(\ell_*,\ell)$,
but because of (\ref{consistency}),
the integral is independent of the path chosen.

The complete effective potential is the
sum of the gluon, quark, and Vandermonde potentials:
\beq
\veff = \calv_{gl} + \calv_{qk} + \veig~.
\eeq
The Vandermonde potential, $\veig(\ell_*,\ell)$,
represents the effects of the $SU(3)$ measure, and so is
invariant under $Z(3)$ transformations,
$\ell\to \exp(2\pi i/3)\ell$ and
$\ell_*\to \exp(-2\pi i/3)\ell_*$.  
In contrast, the quark loop potential is not $Z(3)$ invariant.  

\begin{figure}[htp]
\begin{center}
\epsfxsize=.48\textwidth
\epsfbox{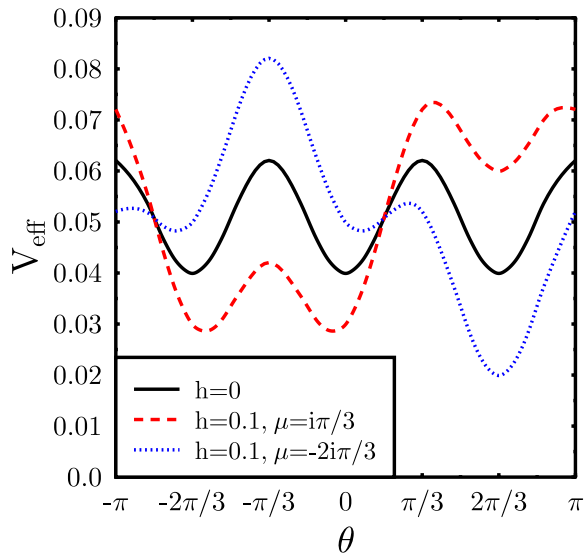}
\end{center}
\vspace*{-.5cm}
\caption{The effective potential $\veff$ for the $SU(3)$ matrix model
  at imaginary $\mu$; there is no gluon loop potential.
  In all curves, $|\ell|=0.2$, with $\theta$ the phase of the loop:
  the solid curve is $h=0$, the dotted curve $h=0.1$ and 
  $\mu = -2 \pi i/3$,
  the dashed curve $h=0.1$ and ${\mu} =  i\pi/3$.
  }
\label{fig:Veff_imagmu}
\end{figure}

As a check on our numerical analysis, 
we first discuss the case where $\mu$ is purely imaginary,
$\mu = i \widetilde{\mu}$, which is a $U(1)$ rotation of the
Wilson line:
\beq
\calv_{qk}(\boldl) = - \frac{h}{2} \left( {\rm e}^{i \widetilde{\mu}} 
\; \ellt \; +
\; {\rm e}^{-i \widetilde{\mu}} \; \ellat \right) \; .
\label{quark_potential_three}
\eeq
If the overall symmetry were $U(3)$, instead of $SU(3)$,
then the Vandermonde potential is independent
of $\widetilde{\mu}$.  For a $SU(3)$ theory, however, 
the $Z(3)$ symmetry only requires that
the potential is degenerate when $\widetilde{\mu} = 0$
and $\pm 2 \pi/3$.

As $\widetilde{\mu}$ represents an ordinary background field,
the anti-triplet loop is the complex conjugate of the triplet loop.
Defining $\theta$ as the phase of $\ell$, 
$\ell = \exp(i \theta) |\ell|$, then $\ell_* = \exp(-i \theta) |\ell|$,
and the Vandermonde potential is a function of $|\ell|$ and $\theta$.

To illustrate the physics, in Fig.~\ref{fig:Veff_imagmu} 
we show three examples, with $h = 0.0$ or $0.1$, and $|\ell| = 0.2$.
When there is no background field, $h=0$,
there are three degenerate minima at $\theta = 0$ and $\pm 2 \pi/3$.
When $h \neq 0$ and
$\widetilde{\mu} = - 2 \pi/3$, the background field
 ``tilts'' the potential so that the expectation value is along
the opposite direction, for $\theta = 2\pi/3$.
Lastly, when $h \neq 0$, and for the special choice of 
$\widetilde{\mu}=\pi/3$, 
the background field points exactly in the direction opposite to the
minimum at $2\pi/3$; then there are two degenerate minima, for
$\theta = 0$ and $-2\pi/3$.  The potential for other values
of $\theta$ and $\ell$ follows similarly. Also, it is clear that,
as a function of $\widetilde{\mu}$, the expectation value of $\theta$ is
discontinuous at $\widetilde{\mu}=\pi/3$, jumping from 0 to
$-2\pi/3$. Analytic continuation of $\calz$ to real $\mu$ is therefore
possible only for $|\mu|<\pi/3$.

\begin{figure}[htp]
\begin{center}
\epsfxsize=.48\textwidth
\epsfbox{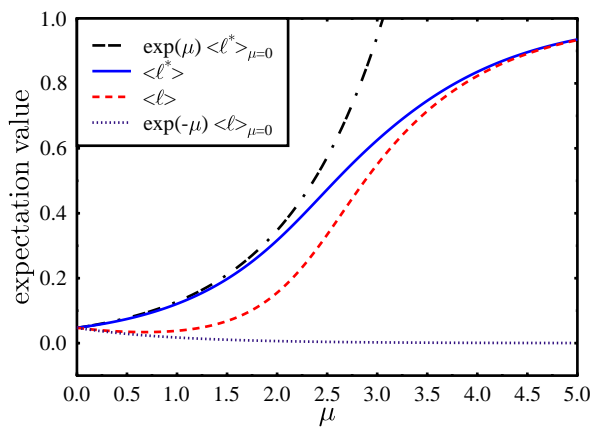}
\end{center}
\vspace*{-.5cm}
\caption{The expectation values $\langle\ell\rangle$ and
$\langle\ell^*\rangle$ as functions of $\mu$ for $h=0.1$ and $m^2=0$.}
\label{fig:vevs_m2=0}
\end{figure}
When the chemical potential is real,
as noted from (\ref{vev_loop}) and (\ref{vev_conj_loop}),
$\langle\ell\rangle$ and
$\langle\ell^* \rangle$ are unequal but real.
Fig.~\ref{fig:vevs_m2=0} shows the expectation values of the loop and
of its conjugate
for a background field $h=0.1$ (again without a gluon loop potential). 
For small $h$ and $\mu=0$,
an analytical discussion of the $N=\infty$ potential at the
Gross-Witten point shows that $\langle\ell\rangle\simeq h/2$, cf.\
section~IIIB in~\cite{dlp}. From Fig.~\ref{fig:vevs_m2=0} one
observes that this remains approximately true also for three colors. 

At non-zero $\mu$ then, $\langle\ell\rangle$ and
$\langle\ell^*\rangle$ split, which is due to the imaginary part of
the fermion contribution~(\ref{qpotentialB}) to the loop action.
While $\langle\ell^*\rangle$ increases monotonically with $\mu$,
$\langle\ell\rangle$ initially {\em decreases} from its value at
$\mu=0$, cf.\ eq.~(\ref{derivatives}).
Finally, both expectation values approach one at large $\mu$, in accord
with our discussion in Sec.~\ref{suN_matrix_model}.

In Sec.~\ref{U1_matrix_model} we saw that in a $U(1)$ model,
the $\mu$-dependence of the expectation values is
entirely given by a factor $\exp(\pm\mu)$, (\ref{equality}).
We have checked numerically that this is approximately valid for
$SU(3)$ when the chemical potential is
very small. Fig.~\ref{fig:vevs_m2=0} shows, however,
that this fails when $\mu\sim 1$.

In a matrix model, $\ell_3$ and $\ell_{\overline{3}}$ are traces
of matrices.  One could also consider a Polyakov loop model
\cite{polyakov_loop}, where $\ell_3$ and $\ell_{\overline{3}}$ are 
just scalar fields.  To reduce the global symmetry
from $U(1)$ to $Z(3)$, it is necessary to include cubic terms,
such as $(\ell_3)^3 + (\ell_{\overline{3}})^3$, (\ref{zna}).  
As for the matrix model, one finds that 
the expectation values of $\ell_3$ and
$\ell_{\overline{3}}$ differ when $\mu \neq 0$.
Their exact form depends upon the details of the loop potential.

We now turn to a discussion of the effective potential
$\veff(\ell,\ell_*)$, for real chemical potential. We also include the
gluon loop potential from eq.~(\ref{gluon_potential}).
The solutions of the stationarity conditions
\beq
\frac{\partial\veff(\ell,\ell_*)}{\partial\ell} =
\frac{\partial\veff(\ell,\ell_*)}{\partial\ell_*} = 0~,
\eeq
determine the expectation values of the triplet and anti-triplet
loops, $\langle\ellt\rangle$ and $\langle\ellat\rangle$.
Due to~(\ref{stationary_leg}), these equations can be rewritten as
\begin{eqnarray}
\omega(\ell^0_*,\ell^0) &=& 
  - \frac{\partial}{\partial\ell}\left[\calv_{gl}(\ell,\ell_*) +
  \calv_{qk}(\ell,\ell_*)\right]
  \Big|_{\ell=\ell^0,\ell_*=\ell^0_*} 
  \nonumber\\
&=& m^2\ell_*^0 + \frac{h}{2}\, e^\mu \label{omega_stat} \\
\omega_*(\ell^0_*,\ell^0) &=& 
  - \frac{\partial}{\partial\ell_*}\left[\calv_{gl}(\ell,\ell_*) +
  \calv_{qk}(\ell,\ell_*)\right]
  \Big|_{\ell=\ell^0,\ell_*=\ell^0_*}\nonumber\\
&=& m^2\ell^0 + \frac{h}{2}\, e^{-\mu}~.  \label{omega*_stat}
\end{eqnarray}
These equations have to be solved simultaneously with~(\ref{lambdas}).
Note that at the stationary point both $\omega$ and $\omega_*$ are
real. Also, these equations,
unlike~(\ref{vev_loop}) and (\ref{vev_conj_loop}) above, make it obvious
that the expectation values of the loops are not determined by
cancellations of positive and negative contributions:
(\ref{omega_stat}) and 
(\ref{omega*_stat}) do not involve any oscillating functions.

To show the shape of the effective potential we
fix $\ell-\ell_*$ to its expectation
value given in eqs.~(\ref{vev_loop},\ref{vev_conj_loop}) or in
eqs.~(\ref{omega_stat},\ref{omega*_stat}) above. We then study $\veff$
as a function of the remaining degree of freedom, $\ell+\ell_*$. 

\begin{figure}[htp]
\begin{center}
\epsfxsize=.48\textwidth
\epsfbox{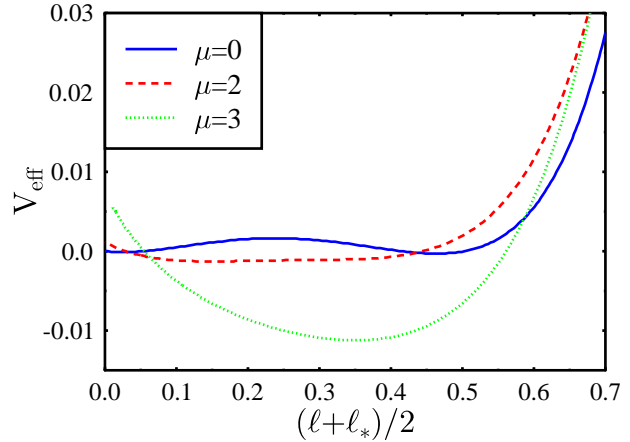}
\end{center}
\vspace*{-.5cm}
\caption{The effective potential at the peak of the Polyakov loop
  susceptibility for $h=10^{-3}$ and various $\mu$.}
\label{fig:veff}
\end{figure}
The behavior of the effective potential with nonzero $h$ and $\mu$
is customary of a first order transition in a background magnetic
field.  
Fig.~\ref{fig:veff} shows the effective potential for $h=10^{-3}$ and 
$\mu=0$, 2, 3, respectively, as a function of $\ell+\ell_*$. For each
curve, the coupling $m^2=m^2_c(\mu,h)$ is
adjusted to maximize the susceptibility $\partial
\langle\ellt + \ellat\rangle/\partial m^2$.
For such a weak background field, the
first-order phase transition persists at $\mu=0$.  As $\mu$ increases,
the two minima of $\veff$ approach each
other and the barrier decreases. The first-order phase transition ends
in a critical point at $\mu=\mu_E$.
The transition is of second order at $\mu_E$,
as the mass of the real part of the triplet loop vanishes.
{}From Fig.~\ref{fig:veff}, $\mu_E \approx 2.0$.
As the chemical potential increases above
$\mu_E$, the mass increases again, and there is no phase transition.

To date, Monte Carlo simulations have been performed for large $T$ and small
$\mu$ \cite{fodor_katz,swansea_bielefeld,deforcrand_philipsen,gavai_gupta}.  
A matrix model predicts that $\langle \ell \rangle \neq
\langle \ell^* \rangle$ when $\mu \neq 0$.  The first work
of Allton {\it et al.} \cite{swansea_bielefeld} did not test 
this directly, but finds that 
$\langle \ell \rangle$ changes when the sign of $\mu$ is flipped.
They did not plot $\langle \ell\rangle$ or $\langle \ell^*\rangle$ versus
$\mu$, and so did not test the prediction that one of these 
expectation values is not monotonic in $\mu$.

The analysis of the present paper is most applicable for heavy
quarks.  At $\mu = 0$, the lattice gives us an outline 
of the phase transition for three degenerate
flavors of massive quarks.
The deconfining transition only persists for relatively
heavy quarks, $m > m_{qk}^{end}$ \cite{lattice_review_temp}; 
in~\cite{DRR}, this 
was estimated to disappear for a pseudo-scalar mass of
$\approx 1.8$~GeV.  There is no phase transition
for intermediate quark masses, with a first order chiral
transition appearing for light quark masses.  In all cases at $\mu = 0$,
the rise in the Polyakov loop appears to coincide with the
decrease in the chiral order parameter.

The case of heavy quarks at $\mu \neq 0$ is then similar to
that of Fig.~\ref{fig:veff}: a first order transition at $\mu = 0$, ending
in a critical end point at some $\mu_E$.  
(See, also, Fig.~1 of~\cite{AKKS}).
For quarks lighter
than $m_{qk}^{end}$, there is no deconfining transition,
and the correlation length of the Polyakov loop decreases monotonically
as $\mu$ increases from zero.  In the plane of $\mu$ and $T$,
there may be a critical end point at $\mu \neq 0$, where the
correlation length for the sigma meson diverges \cite{srs}; that for
the Polyakov loop will remain finite, except from its coupling
to the sigma.

To describe the region of small 
quark masses, and the chiral transition, it is necessary
to introduce a chiral order parameter, and couple that to the
Wilson line.  A mean field approximation can be analyzed in
a matrix model with two coupled matrices \cite{toappear}.
Due to the large $N$ argument
mentioned at the end of Sec.~II, it is possible that for three
colors, the coincidence of the chiral and deconfining ``transitions'',
ubiquitous at $\mu = 0$, breaks down at some finite value of $\mu$.
That is, for intermediate quark densities, hadronic matter exists
as a Fermi sea of ``confined'', but chirally symmetric, nucleons.

\section{Lattice QCD and the Fermion Sign Problem}
\label{Lattice_QCD}

We conclude by discussing how the results of the matrix model may 
be of use
for numerical simulations of dense quarks in lattice QCD.

In Euclidean spacetime, the quark part of the action is 
\beq
{\cal S}_{qk} \; = \; \int d^4x \; \; \overline{\psi}
\left( \dslash + \mu \gamma^0 + m \right) \psi \; .
\label{quark_action}
\eeq
We follow the conventions of \cite{zinn_justin}, with
$\dslash = (\partial_\mu - i g A_\mu)\gamma^\mu$ the
covariant derivative for a gluon field $A_\mu$.  
In this section, and in contrast to previous notation, here $\mu$ is
the quark chemical potential (not $\mu/T$), and
$m$ is the quark mass (not $m_{qk}$).

We need to use two symmetries.  By a combination
of Hermitian conjugation, plus a $\gamma_5$ transformation,
\beq
\left(\detrm (\dslash + \mu \gamma^0 + m )\right)^* \; = \;
\detrm(\dslash - \mu \gamma^0 + m ) \; ;
\label{complex_quark_op}
\eeq
see, {\it e.g.}, (13) of \cite{splittorff1}.  
At zero chemical potential, $\dslash$ is purely anti-Hermitian;
as $\dslash$ anti-commutes with the matrix $\gamma_5$, the eigenvalues
pair up, and the quark determinant is real.  At nonzero chemical
potential, the quark operator is a sum of an anti-Hermitian operator,
$\dslash$, and a Hermitian operator, $\mu \gamma_0$.  While
the eigenvalues form pairs with opposite sign,
(\ref{complex_quark_op}) shows that the quark determinant is complex
when $\mu \neq 0$.  This is the fermion sign problem in dense QCD.

We can perform a charge conjugation transformation on the quarks
\cite{zinn_justin}.  This is a change of variables in the Grassman
integration over the quarks, and so it doesn't change
the determinant.  This gives
\beq
\detrm\left(\dslash + \mu \gamma^0 + m \right) \; = \;
\detrm\left(\dslash_c - \mu \gamma^0 + m \right) \; ,
\label{charge_conj}
\eeq
where $\dslash_c = (\partial_\mu + i g A_\mu^*)\gamma^\mu$
is the covariant derivative for the charge conjugate gluon
field, $- A_\mu^*$.  By itself, this isn't of much help, as
we have changed the sign of the chemical potential, and turned
the gluon field into its charge conjugate.  In the matrix model,
this symmetry is manifest in (\ref{quark_potential}).

We now combine these two relations, to obtain:
\beq
\detrm(\dslash_c + \mu \gamma^0 + m) \; = \;
\left(\detrm (\dslash + \mu \gamma^0 + m )\right)^* \; .
\label{charge_conj_det}
\eeq
This shows that for the same sign of $\mu$, the quark determinant
for charge conjugate gluons is the complex conjugate of the quark
determinant in the original gluon field.  This generalizes
what is obvious in the matrix model.

Thus while the quark determinant in the presence of a given
gluon field is complex, by adding the contribution of the charge
conjugate gluons, we immediately obtain a partition function whose
measure is manifestly real.  This extends immediately to the lattice.
There, gluons live on links, with link fields $U_\mu
= \exp(i g a A_\mu)$, where ``$a$'' is the lattice spacing.  
A configuration of links is given by some set of $U_\mu$'s;
the charge conjugate lattice is simply given by replacing
each $U_\mu$ by $U_\mu^*$.  That one can, in this way, 
obtain a real measure of the functional integral 
was known from the work of the Glasgow group 
(see, {\it e.g.}, the discussion just before 
Eq. (8) in the last reference of \cite{glasgow}).

However, all this does is to reduce the problem from one of complex
phases, to one of real phases.  There are still configurations in
the functional integral with both positive and negative weight.
This still leaves the problem of how to decide whether to sum
over configurations with both signs.  Also, how does one include the
effects of a Fermi sea of quarks in weighting configurations?

The matrix model provides clues to both of these questions.  
It is true that configurations of both signs contribute to the integral of 
the matrix model.  However, at zero density, the background \zn field
which quarks induce provides an expectation value along a definite
direction in the complex plane, for real values (this is related to
the sign of the quark masses).  Further, at nonzero quark density,
the field for the imaginary part of the loop has an imaginary coefficient,
so that the expectation values of both $\ell$ and $\ell^*$ remain
real and positive.  
We have checked that even at nonzero $\mu$, the dominant
configurations of the matrix model are those in which the measure
is {\it positive}.  

It is reasonable to conjecture that this remains
true with dynamical quarks.  This suggests that in Monte Carlo simulations,
that one accept configurations in which the quark determinant is positive,
and drop those in which it is negative.

How, then, does one weight by configurations which include the
effects of a Fermi sea of quarks?  
This could be included by a type of tadpole improvement.  Suppose
that one works out from zero chemical potential, to increasingly
large values.  To represent the effects of $\mu \neq 0$, one would
expand not about the bare link variables, but about links equal
to the expectation value of the loop.  For a link going forward,
one would use $\langle \ell\rangle$; for a link going backward,
$\langle \ell^*\rangle$.  This will explicitly bias one to configurations
which include, approximately, the effects of the Fermi sea.

This is supported by numerical simulations of Blum, Hetrick,
and Toussaint for heavy quarks \cite{blum}.
Using their results, de Forcrand and Laliena \cite{deforcrand_lal}
showed that the phase of the (heavy) quark
determinant is proportional to the phase of the Polyakov loop,
times the spatial volume.

These results illustrate a more general problem.
The parameters of a matrix model are just numbers.  This
represents, however, a mean field approximation to the theory 
in a spatial volume, $V$.  For example, 
the background field induced by
quarks is itself proportional
to $V$.  For a measure which is always real and positive,
this is of no concern: even if $h \sim V$, 
an error of order one is inessential relative to the
dominant term, which is $\exp(-\# h) \sim \exp(-\# V)$.  
The integrals which enter at nonzero density, however, are those
in which the measure includes oscillatory terms, as in
(\ref{partition_func_real}),
(\ref{vev_loop}), and (\ref{vev_conj_loop}).  In this case,
it is necessary to determine the phase accurately not just to
$\sim V$, but to $\sim 1$!  In essence, this is the true fermion
sign problem: not that the measure is not positive definite, but
that one must determine the phase of the quark determinant {\it very}
accurately.  We note that similar oscillations in the quark determinant
have been derived, using random matrix theory in the $\epsilon$-regime,
by Osborn, Splittorff and Verbaarschot \cite{splittorff2}.

Nevertheless, we suggest that these techniques might be of use
in numerical simulations of dense QCD on the lattice.  By their
nature, they are most suited for heavy quarks, starting from the
region of zero density, and working out to nonzero density.  
Even if one accepts configurations whose overall weight is positive,
it is certainly necessary to use cluster algorithms to include
regions in which the phase is negative \cite{merons_etc}.  

We have used an effective model which is, implicitly, valid only
for distances $\gg 1/T$.  When the temperature is small, 
it is also imperative to include fluctuations in
the expectation values of timelike links, as they wander about in (imaginary)
time.  

Lastly, these ideas are strongly motivated by the heavy quark
limit \cite{blum,deforcrand_lal}, 
where quarks only propagate upward in imaginary time.  
Light quarks also 
propagate in space, so that at nonzero density, 
one will have to expand about modified expectation values for
propagation which is ``forward'' or ``backward'' in proper time.

There is now a wealth of results available at nonzero temperature
and small chemical potential 
\cite{fodor_katz,swansea_bielefeld,deforcrand_philipsen,gavai_gupta,splittorff1,splittorff2}.  This is the first place to test our 
admittedly speculative remarks.

{\bf Acknowledgements:}
R.D.P.\ is supported by the U.S.\ Department of Energy grant DE-AC02-98CH10886,
and thanks the Alexander von Humboldt Foundation for their support.
He also thanks the following for discussions:
T.\ Blum, M.\ Creutz, Dr.\ Flotte, K. Fukushima, T.\ Izubuchi, F.\ Karsch, 
S.\ Ohta,
P.\ Petreczky, K.\ Splittorff, and most especially, D.\ H.\ Rischke.

\end{document}